\documentclass[aps,prl,floatfix,superscriptaddress,showpacs,twocolumn]{revtex4}
\usepackage{amsmath}
\usepackage{graphicx}

\begin{document}
\title{Hyperpolarized xenon nuclear spins detected by optical atomic magnetometry}
\author{V. V. Yashchuk}
\email{yashchuk@socrates.berkeley.edu} \affiliation{Department of
Physics, University of California at Berkeley, Berkeley,
California 94720-7300}
\author{J. Granwehr}\email{joga@waugh.cchem.berkeley.edu}
\affiliation{Materials Sciences Division, Lawrence Berkeley
National Laboratory, and Department of Chemistry, University of
California at Berkeley, Berkeley, California 94720-1460}
\author{D. F. Kimball}
\email{dfk@uclink4.berkeley.edu} \affiliation{Department of
Physics, University of California at Berkeley, Berkeley,
California 94720-7300}
\author{S. M. Rochester}
\email{simonkeys@yahoo.com} \affiliation{Department of Physics,
University of California at Berkeley, Berkeley, California
94720-7300}
\author{A. H. Trabesinger}
\email{Andreas.Trabesinger@nmr.phys.chem.ethz.ch}
\affiliation{Laboratory of Physical Chemistry ETH-H\"{o}nggerberg,
CH-8093 Zurich, Switzerland}
\author{J. T. Urban}\email{jurban@OCF.Berkeley.EDU}
\affiliation{Materials Sciences Division, Lawrence Berkeley
National Laboratory, and Department of Chemistry, University of
California at Berkeley, Berkeley, California 94720-1460}
\author{D. Budker}
\email{budker@socrates.berkeley.edu} \affiliation{Department of
Physics, University of California at Berkeley, Berkeley,
California 94720-7300} \affiliation{Nuclear Science Division,
Lawrence Berkeley National Laboratory, Berkeley, California 94720}
\author{A. Pines}\email{pines@berkeley.edu}
\affiliation{Materials Sciences Division, Lawrence Berkeley
National Laboratory, and Department of Chemistry, University of
California at Berkeley, Berkeley, California 94720-1460}

\date{\today}
\begin{abstract}
We report the use of an atomic magnetometer based on nonlinear
magneto-optical rotation with frequency modulated light (FM NMOR)
to detect nuclear magnetization of xenon gas. The magnetization of
a spin-exchange-polarized xenon sample ($1.7\ $cm$^3$ at a
pressure of $5\ $bar, natural isotopic abundance, polarization $1\
\%$), prepared remotely to the detection apparatus, is measured
with an atomic sensor (which is insensitive to the leading field
of 0.45 G applied to the sample; an independent bias field at the
sensor is $140\ \mu$G). An average magnetic field of $\sim 10\ $nG
induced by the xenon sample on the 10-cm diameter atomic sensor is
detected with signal-to-noise ratio $\sim 10$, limited by residual
noise in the magnetic environment. The possibility of using modern
atomic magnetometers as detectors of nuclear magnetic resonance
and in magnetic resonance imaging is discussed. Atomic
magnetometers appear to be ideally suited for emerging low-field
and remote-detection magnetic resonance applications.
\end{abstract}



\maketitle

Nuclear magnetic resonance (NMR) is a versatile technique for the
study of structure and dynamics on both molecular and macroscopic
scales, and on time scales from nanoseconds to hours.
Spin-polarized $^{129}$Xe (nuclear spin-1/2, magnetic moment
$\mu\approx-0.78\ \mu_N$, where $\mu_N$ is the nuclear magneton)
is particularly well suited for NMR and magnetic-resonance imaging
(MRI) studies for several reasons. It is possible to polarize it
using the laser-optical-pumping techniques \cite{Wal97} to a
degree that is orders of magnitude higher than what is possible
via thermal polarization in high-field magnets. In contact with
various analytes, xenon displays a wide range of relative chemical
shifts of up to several hundred ppm \cite{Goo2002}, which makes it
an ideal probe of its local physiochemical environment. Finally,
it has a long longitudinal relaxation time of several minutes or
longer, even at low fields. Xenon can also be used in solution,
and is especially soluble in organic solvents.

An important recent development in NMR/MRI is the technique of
remote detection \cite{Mou2003,See2004}, in which information
about an analyte is transferred onto a mobile spin-polarized
substance, and is then read out at a different location. This
technique allows the separate optimization of the encoding and the
detection environment. As the signal in this case is due to the
net magnetization of the spin-polarized sample, the task becomes
to read out this information efficiently and with high
sensitivity. Detection using superconducting quantum interference
devices (SQUIDs) \cite{Cla96} and atomic magnetometers
\cite{Bud2002RMP} provides an alternative to the traditional
techniques involving inductive detection. In addition to
eliminating the need for a strong magnetic field for the detector,
another advantage of these methods is that the time constant of
the measurement in this case is limited by the longitudinal
relaxation, which could be significantly slower than transverse
relaxation limiting induction detection. SQUID detection has
already proven useful in NMR experiments \cite{Gre98,McD2002}.

Atomic magnetometers (the essential components of which consist
only of a diode laser, an atomic vapor cell, and the necessary
optics and electronics) are also attractive for NMR applications
as they have the potential to be very cheap and compact and---in
contrast to SQUIDs, which require cryogenic temperatures---they
operate at room temperature. The first use of an atomic
magnetometer for detection of the static magnetic field produced
by a sample of gaseous nuclear-polarized atoms was reported nearly
35 years ago in the pioneering work of Cohen-Tannoudji \textit{et
al.} \cite{Coh69}. In that work, a 6-cm diameter vapor cell
containing 5$\%$-optically-polarized $^3$He gas at a pressure of 3
torr was placed next to a cell of similar dimensions containing
$^{87}$Rb that served as a sensor of an optical-pumping
magnetometer. The $60~{\rm nG}$ field produced by the nuclear
spins was detected with a sensitivity $\sim 3\times 10^{-9}\
\rm{G}/\sqrt{\rm{Hz}}$. A similar setup was also used in Ref.
\cite{New93}.

Modern atomic magnetometers utilizing alkali-metal vapors in
anti-relaxation coated cells can achieve sensitivities to magnetic
field better than $10^{-11}\ $G$/\sqrt{\rm{Hz}}$, see, for
example, Ref. \cite{Bud2000Sens} and review \cite{Bud2002RMP}.
Recently, a sensitivity of $5\times 10^{-12}\
\rm{G}/\sqrt{\rm{Hz}}$ was demonstrated \cite{Kom2003} with an
atomic sensor of a volume of only $0.3\ $cm$^3$ where, instead of
anti-relaxation coating, buffer gas was used to reduce relaxation
in wall collisions, and operation at a high alkali-atom density
ensured the rapid spin-exchange-collision regime \cite{Hap77},
where spin-relaxation in collisions between alkali atoms is also
reduced. Due to such advances, the use of atomic magnetometers for
low-field NMR experiments becomes attractive.

\begin{figure}
\bigskip
\includegraphics[width=3.5 in]{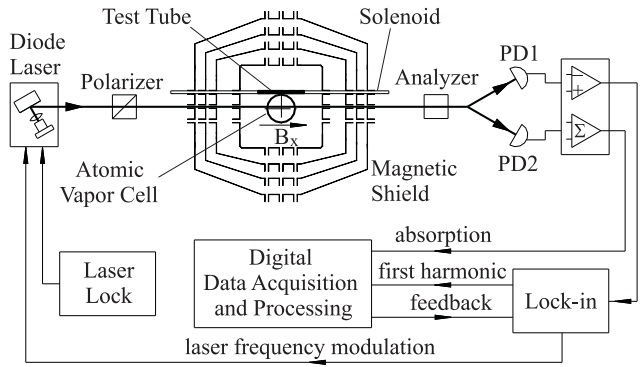}
\caption{Atomic magnetometer used for detecting Xe nuclear-spin
polarization.}\label{FigXeSetUp}
\end{figure}

\begin{figure}
    \bigskip
    \includegraphics[width=3.25 in]{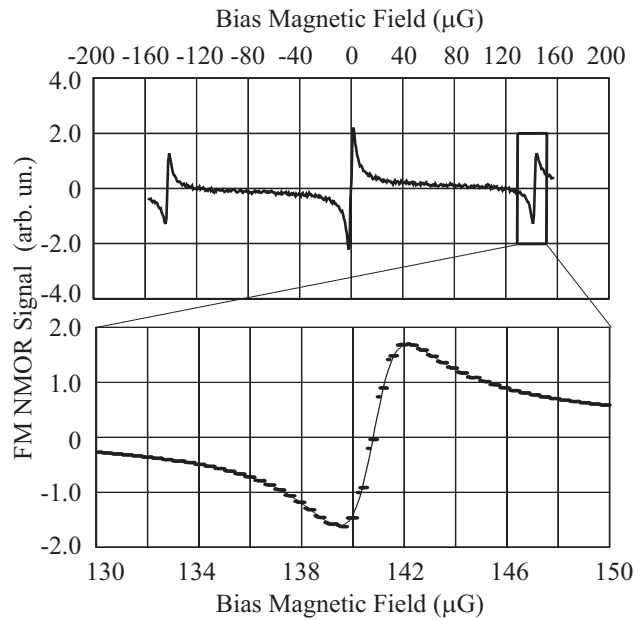}
    \caption{The FM NMOR resonances recorded with laser-modulation
    frequency set at 200$\ $Hz. During the operation of the
    magnetometer, the bias field is fixed and the modulation frequency
    is locked to the center of a resonance.}\label{FigBxDependence}
\end{figure}

With a view toward future NMR applications, particularly remote
detection, we have carried out exploratory measurements of samples
of gaseous spin-exchange-polarized xenon with a modified version
of the atomic magnetometry apparatus previously described in
Refs.\ \cite{Bud2000Sens,Bud2002FM,Mal2004} (Fig.\
\ref{FigXeSetUp}).
The apparatus incorporates a 10-cm diameter spherical
$^{87}$Rb-vapor cell at room temperature with no buffer gas, whose
inner walls are coated \cite{AleLIAD} with paraffin to reduce spin
relaxation in collisions of Rb atoms with the wall. The
magnetic-field measurement is based on the technique of nonlinear
magneto-optical rotation with frequency-modulated light (FM NMOR)
\cite{Bud2002FM}. The key feature of the method is the use of an
ultra-narrow resonance arising when the frequency of the light is
modulated at twice the Larmor-precession frequency of the Rb atoms
(Fig. \ref{FigBxDependence}). The magnetometer operates in a
closed feedback loop involving digital signal processing, locking
to an FM NMOR resonance by adjusting the diode-laser-modulation
frequency. The mean frequency of the $4\ \mu$W laser light
delivered to the sensor is locked to the D1-resonance using a
technique \cite{Yas2004ModDAVLL} involving an auxiliary Rb-vapor
cell (not shown). The atomic sensor cell is placed inside a
multi-layer magnetic shield (shielding factor $\sim10^6$
\cite{Yas99}) equipped with internal magnetic-field coils. The FM
NMOR magnetometer is intrinsically a scalar device; however, a
bias field $B_x=140\ \mu$G applied along the direction of light
propagation renders the sensor linearly sensitive only to the
average component of the magnetic field produced by the sample in
that direction.

The xenon samples of natural isotopic abundance containing  about
26$\%$ of $^{129}$Xe were prepared using a commercial
spin-exchange polarizer [MITI IGI 9800 Xe, Magnetic Imaging
Technologies, Inc. (Polarean), Durham, NC]. A gas mixture of 1\%
Xe, 10\% N$_2$, and 89\% He was used. After polarization, pure
xenon was frozen out of the gas mixture into a cold finger that
was immersed in liquid nitrogen. At the end of the polarization
process, the Xe batch was thawed into a custom sapphire sample
tube equipped with a miniature titanium spring-loaded valve
mechanism with overall outer diameter of 6.4$\ $mm and sample
volume dimensions of 4.8$\ $mm inner diameter and $140\ $mm length
\cite{Yas2004NMRtube}. These materials were chosen to ensure a
long spin-polarization relaxation time \cite{Haa98}. The outer
diameter of the tube was dictated by the tight constraints given
by the geometry of the atomic magnetometer's magnetic shield
\cite{Yas99}. The tube was designed for operation with gas
pressures of up to 30$\ $bar. The xenon polarization was measured
by the polarizer's onboard NMR spectrometer and calibrated using
thermally-polarized xenon dissolved in pentane on a Varian Unity
Inova NMR spectrometer equipped with an Oxford 7-T magnet. The
longitudinal relaxation time constant $T_1$ in the tube was found
to be $T_1\approx 45\ $min within the 7-T magnet, and typically
$\lesssim 15\ $min in the earth and laboratory fields. For the
experimental runs described here, we used samples with total xenon
pressure of $\sim 5\ $bar with $4-8\%$ initial $^{129}$Xe
polarization.

\begin{figure}
\bigskip
\includegraphics[width=3.5 in]{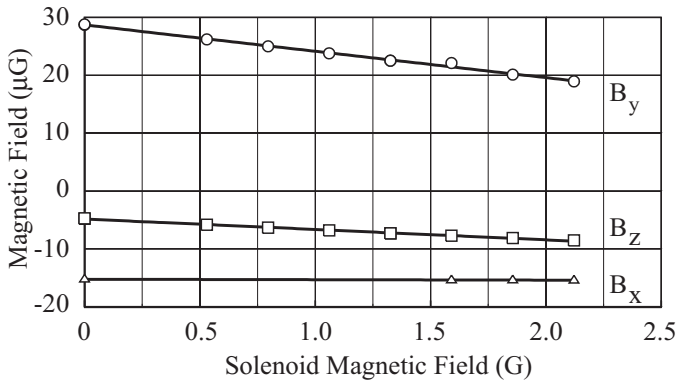}
\caption{Auxiliary measurements of the field leaking from the
leading-field solenoid. These measurements were performed at the
zero-field FM NMOR resonance (Fig. \ref{FigBxDependence}), that is
sensitive to transverse components of the field. During xenon
measurements, the residual fields were nulled with the internal
coils.}\label{FigBxVsCurrent.eps}
\end{figure}

A piercing solenoid (see Fig. \ref{FigXeSetUp}) was used to apply
a leading field of 0.45 G to the xenon sample. The setup is
designed so that the leakage field outside of the solenoid has
negligible effect on the atomic sensor; the field due to the
piercing solenoid is a factor of $\approx 2 \times 10^5$ smaller
at the vapor cell, as determined by an auxiliary measurement with
the atomic magnetometer (Fig. \ref{FigBxVsCurrent.eps}). During
the xenon measurement, this leakage field provides a constant
magnetic field offset that is much smaller than and transverse to
the $140\ \mu$G bias field applied to the atomic sensor. Thus, the
FM NMOR measurement is largely insensitive to the solenoid leakage
field.

\begin{figure}
\bigskip
\includegraphics[width=3.25 in]{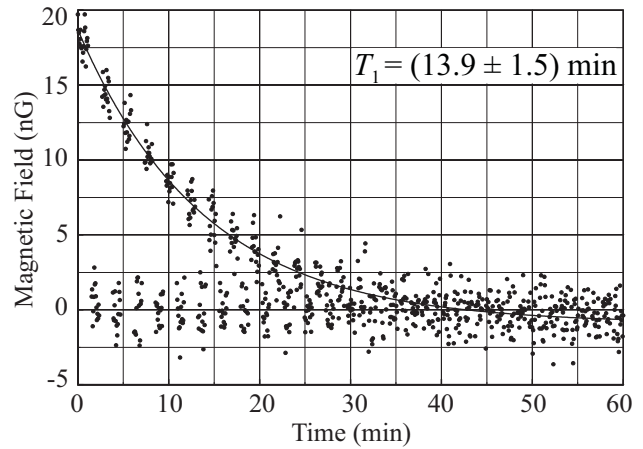}
\caption{An example of the Xe-magnetization signal (the average
magnetic field produced by the xenon sample over the rubidium
sensor cell) recorded with the atomic magnetometer. Approximately
every minute, the xenon sample is moved within the solenoid
between the position of maximum and near-zero sensitivity (when
the sample is partially outside of the innermost magnetic shield)
of the atomic magnetometer to the magnetic field produced by the
sample. We have subtracted from the data the overall slow drift
due to the temperature drift-related change of the residual
magnetization of the shield. This drift is determined from the
portion of the data taken when the xenon sample is in the position
where the sensor does not feel its magnetization. There is a
slight offset of the data ($\lesssim 1\ $nG) due to the
diamagnetic susceptibility of the sample tube.}\label{FigXeData}
\end{figure}

After introducing spin-exchange-polarized xenon into the tube, the
sample is hand-delivered to the atomic magnetometry laboratory
located in a different building and loaded into the magnetometer
apparatus inside the piercing solenoid (Fig. \ref{FigXeSetUp}).
Typically, it takes about 5$\ $min between the completion of xenon
gas preparation and the beginning of the atomic magnetometer
measurements. Differential measurement is achieved by moving the
sample within the piercing solenoid in and out of the position
next to the Rb cell where maximal sensitivity to the xenon
magnetic field is obtained. This modulation is effective in
discriminating between the xenon magnetization signal and the slow
drift (typically, $\sim 1\ $nG/min) of the magnetic field within
the shield. When the sample tube is in the position where the
atomic sensor is most sensitive to its field, the average magnetic
field from the xenon magnetization over the volume of the
magnetometer sensor cell (equal to the field in the center of the
cell) is a factor of 5000 smaller than the field in the sample
tube. This suppression factor was calculated from the experimental
geometry and verified experimentally by replacing the sample cell
with a calibrated solenoid of the same dimensions, and could be
greatly reduced with a sensor geometry designed specifically for
this application. When the sample tube is moved away to the ``out"
position, the suppression factor is about two orders of magnitude
larger. An example of experimental data is shown in Fig.
\ref{FigXeData}.

With a measurement time of about 3$\ $s per point, we detect the
decay of xenon magnetization corresponding to an initial average
field at the atomic sensor of $10-20\ $nG with a signal-to-noise
ratio (S/N) of about 10, clearly demonstrating the ability of the
atomic sensor to detect dc magnetization of a small gaseous
sample. It should be emphasized that the S/N obtained in this work
is many orders of magnitude lower than what one could obtain with
straightforward modifications of this technique. Perhaps most
significant would be the improvement of the geometrical
suppression factor from 5000 to about 10 with optimized geometry.
In addition, the magnetic noise in this experiment is dominated by
the fluctuations of the magnetic field within the magnetic shield
and exceeds the projected intrinsic sensor noise of our apparatus
\cite{Bud2000Sens,Bud2002FM} of $\lesssim 10^{-11}\
$G$/\sqrt{\rm{Hz}}$ by about two orders of magnitude. This noise
can be effectively suppressed (as was done, for example, in Ref.
\cite{Kom2003}) by employing a gradiometric arrangement of
magnetic sensors.

With a system consisting of two compact, high-precision atomic
magnetometers operating in gradiometric mode, it should be
possible to measure the magnetization of about $10^{13}$
fully-polarized nuclear (e.g., $^{129}$Xe) spins in less than a
second with a signal-to-noise ratio of 10. The open geometry of an
atomic magnetometer equipped with a piercing solenoid would allow
measurements in which polarized samples can be continuously
transported through the magnetometer, an important feature for
remote detection experiments \cite{Mou2003}. Commercial xenon
hyperpolarization systems are capable of producing over a liter of
xenon gas at 1$\ $bar with typical polarization of $\sim 8\%$.
Using such a system and an optimized atomic magnetometer, it will
be possible to make point-by-point low-field measurements, in
which a single-point sample will constitute $\sim 0.1\ $mL of
xenon gas whose magnetization will be determined with S/N
$\sim10^4$ in $\sim 0.3\ $s, allowing ten thousand single-point
measurements to be taken in less than an hour. Faster and/or
higher resolution scans can, in principle, be obtained by using
multiple atomic sensors in parallel, and/or by sacrificing the
S/N.

In principle, it is also possible to perform manipulations on the
nuclear spins within the magnetometer, including adiabatic
spin-flips, spin-echoes, etc. This capability may be of interest
for ultra-low field NMR studies (leading fields can be as low as
$10^{-7}\ $G limited by the magnetic-shielding system; if
necessary, much larger leading fields than those used in this work
may be applied as well). Ultra-low field NMR with SQUID detection
has been recently demonstrated as a powerful tool for analytical
chemistry \cite{McD2002}. A useful feature of the present approach
is the ability, due to the presence of the piercing solenoid, to
apply a leading field to the sample under investigation and an
independent bias field to the atomic-magnetometer sensor cell.

In conclusion, we have demonstrated reliable detection of nuclear
spin polarization of gaseous samples of spin-exchange-polarized
xenon using an atomic magnetometer based on nonlinear
magneto-optical rotation with frequency-modulated light. The
present apparatus is not optimized for NMR/MRI work, and there is
a large ($\sim 5000$) geometrical suppression factor that will be
reduced in a future dedicated setup. That setup will also employ a
gradiometric arrangement of atomic sensors, thus reducing the
presently dominant source of noise resulting from fluctuations and
drift of the magnetic field within the magnetic shield. Estimates
of the sensitivity of such a device show its great promise as a
detector for NMR and MRI studies.

The authors are grateful to Dr. Song-I Han for useful discussions
and to A. Vaynberg, T. Millet, and M. Solarz for making crucial
parts of the apparatus and technical assistance. This work was
supported by the Office of Naval Research (grant
N00014-97-1-0214), by NSF, and by the Director, Office of Science,
Office of Basic Energy Sciences, Materials Sciences and Nuclear
Science Divisions, of the U.S. Department of Energy under contract
DE-AC03-76SF00098. J.G. gratefully acknowledges the Swiss National
Science Foundation for support through a postdoctoral fellowship.

\bibliography{NQR_NMR}

\begin{thebibliography}{20}
\expandafter\ifx\csname natexlab\endcsname\relax\def\natexlab#1{#1}\fi
\expandafter\ifx\csname bibnamefont\endcsname\relax
  \def\bibnamefont#1{#1}\fi
\expandafter\ifx\csname bibfnamefont\endcsname\relax
  \def\bibfnamefont#1{#1}\fi
\expandafter\ifx\csname citenamefont\endcsname\relax
  \def\citenamefont#1{#1}\fi
\expandafter\ifx\csname url\endcsname\relax
  \def\url#1{\texttt{#1}}\fi
\expandafter\ifx\csname urlprefix\endcsname\relax\def\urlprefix{URL }\fi
\providecommand{\bibinfo}[2]{#2}
\providecommand{\eprint}[2][]{\url{#2}}

\bibitem[{\citenamefont{Walker and Happer}(1997)}]{Wal97}
\bibinfo{author}{\bibfnamefont{T.~G.} \bibnamefont{Walker}} \bibnamefont{and}
  \bibinfo{author}{\bibfnamefont{W.}~\bibnamefont{Happer}},
  \bibinfo{journal}{Rev. Mod. Phys.} \textbf{\bibinfo{volume}{69}},
  \bibinfo{pages}{629} (\bibinfo{year}{1997}).

\bibitem[{\citenamefont{Goodson}(2002)}]{Goo2002}
\bibinfo{author}{\bibfnamefont{B.~M.} \bibnamefont{Goodson}},
  \bibinfo{journal}{J. Mag. Reson.} \textbf{\bibinfo{volume}{155}},
  \bibinfo{pages}{157} (\bibinfo{year}{2002}).

\bibitem[{\citenamefont{Moul\'{e} et~al.}(2003)\citenamefont{Moul\'{e}, Spence,
  Han, Seeley, Pierce, Saxena, and Pines}}]{Mou2003}
\bibinfo{author}{\bibfnamefont{A.~J.} \bibnamefont{Moul\'{e}}},
  \bibinfo{author}{\bibfnamefont{M.~M.} \bibnamefont{Spence}},
  \bibinfo{author}{\bibfnamefont{S.}~\bibnamefont{Han}},
  \bibinfo{author}{\bibfnamefont{J.~A.} \bibnamefont{Seeley}},
  \bibinfo{author}{\bibfnamefont{K.~L.} \bibnamefont{Pierce}},
  \bibinfo{author}{\bibfnamefont{S.}~\bibnamefont{Saxena}}, \bibnamefont{and}
  \bibinfo{author}{\bibfnamefont{A.}~\bibnamefont{Pines}},
  \bibinfo{journal}{Proc. Natl. Acad. Sci. U. S. A}
  \textbf{\bibinfo{volume}{100}}, \bibinfo{pages}{9122} (\bibinfo{year}{2003}).

\bibitem[{\citenamefont{Seeley et~al.}(2004)\citenamefont{Seeley, Han, and
  Pines}}]{See2004}
\bibinfo{author}{\bibfnamefont{J.~A.} \bibnamefont{Seeley}},
  \bibinfo{author}{\bibfnamefont{S.}~\bibnamefont{Han}}, \bibnamefont{and}
  \bibinfo{author}{\bibfnamefont{A.}~\bibnamefont{Pines}}, \bibinfo{journal}{J.
  Mag. Reson.} \textbf{\bibinfo{volume}{167}}, \bibinfo{pages}{282}
  (\bibinfo{year}{2004}).

\bibitem[{\citenamefont{Clarke}(1996)}]{Cla96}
\bibinfo{author}{\bibfnamefont{J.}~\bibnamefont{Clarke}}, in
  \emph{\bibinfo{booktitle}{SQUID Sensors: Fundamentals, Fabrication, and
  Applications}}, edited by
  \bibinfo{editor}{\bibfnamefont{H.}~\bibnamefont{Weinstock}}
  (\bibinfo{publisher}{Kluwer Academic}, \bibinfo{address}{The Netherlands},
  \bibinfo{year}{1996}), pp. \bibinfo{pages}{1--62}.

\bibitem[{\citenamefont{Budker et~al.}(2002{\natexlab{a}})\citenamefont{Budker,
  Gawlik, Kimball, Rochester, Yashchuk, and Weis}}]{Bud2002RMP}
\bibinfo{author}{\bibfnamefont{D.}~\bibnamefont{Budker}},
  \bibinfo{author}{\bibfnamefont{W.}~\bibnamefont{Gawlik}},
  \bibinfo{author}{\bibfnamefont{D.~F.} \bibnamefont{Kimball}},
  \bibinfo{author}{\bibfnamefont{S.~M.} \bibnamefont{Rochester}},
  \bibinfo{author}{\bibfnamefont{V.~V.} \bibnamefont{Yashchuk}},
  \bibnamefont{and} \bibinfo{author}{\bibfnamefont{A.}~\bibnamefont{Weis}},
  \bibinfo{journal}{Rev. Mod. Phys.} \textbf{\bibinfo{volume}{74}},
  \bibinfo{pages}{1153} (\bibinfo{year}{2002}{\natexlab{a}}).

\bibitem[{\citenamefont{McDermott et~al.}(2002)\citenamefont{McDermott,
  Trabesinger, Muck, Hahn, Pines, and Clarke}}]{McD2002}
\bibinfo{author}{\bibfnamefont{R.}~\bibnamefont{McDermott}},
  \bibinfo{author}{\bibfnamefont{A.~H.} \bibnamefont{Trabesinger}},
  \bibinfo{author}{\bibfnamefont{M.}~\bibnamefont{Muck}},
  \bibinfo{author}{\bibfnamefont{E.~L.} \bibnamefont{Hahn}},
  \bibinfo{author}{\bibfnamefont{A.}~\bibnamefont{Pines}}, \bibnamefont{and}
  \bibinfo{author}{\bibfnamefont{J.}~\bibnamefont{Clarke}},
  \bibinfo{journal}{Science} \textbf{\bibinfo{volume}{295}},
  \bibinfo{pages}{2247} (\bibinfo{year}{2002}).

\bibitem[{\citenamefont{Greenberg}(1998)}]{Gre98}
\bibinfo{author}{\bibfnamefont{Y.~S.} \bibnamefont{Greenberg}},
  \bibinfo{journal}{Rev. Mod. Phys.} \textbf{\bibinfo{volume}{70}},
  \bibinfo{pages}{175} (\bibinfo{year}{1998}).

\bibitem[{\citenamefont{Cohen-Tannoudji
  et~al.}(1969)\citenamefont{Cohen-Tannoudji, DuPont-Roc, Haroche, and
  Lalo\"{e}}}]{Coh69}
\bibinfo{author}{\bibfnamefont{C.}~\bibnamefont{Cohen-Tannoudji}},
  \bibinfo{author}{\bibfnamefont{J.}~\bibnamefont{DuPont-Roc}},
  \bibinfo{author}{\bibfnamefont{S.}~\bibnamefont{Haroche}}, \bibnamefont{and}
  \bibinfo{author}{\bibfnamefont{F.}~\bibnamefont{Lalo\"{e}}},
  \bibinfo{journal}{Phys. Rev. Lett.} \textbf{\bibinfo{volume}{22}},
  \bibinfo{pages}{758} (\bibinfo{year}{1969}).

\bibitem[{\citenamefont{Newbury et~al.}(1993)\citenamefont{Newbury, Barton,
  Bogorad, Cates, Gatzke, Mabuchi, and Saam}}]{New93}
\bibinfo{author}{\bibfnamefont{N.~R.} \bibnamefont{Newbury}},
  \bibinfo{author}{\bibfnamefont{A.~S.} \bibnamefont{Barton}},
  \bibinfo{author}{\bibfnamefont{P.}~\bibnamefont{Bogorad}},
  \bibinfo{author}{\bibfnamefont{G.~D.} \bibnamefont{Cates}},
  \bibinfo{author}{\bibfnamefont{M.}~\bibnamefont{Gatzke}},
  \bibinfo{author}{\bibfnamefont{H.}~\bibnamefont{Mabuchi}}, \bibnamefont{and}
  \bibinfo{author}{\bibfnamefont{B.}~\bibnamefont{Saam}},
  \bibinfo{journal}{Phys. Rev. A} \textbf{\bibinfo{volume}{48}},
  \bibinfo{pages}{558} (\bibinfo{year}{1993}).

\bibitem[{\citenamefont{Budker et~al.}(2000)\citenamefont{Budker, Kimball,
  Rochester, Yashchuk, and Zolotorev}}]{Bud2000Sens}
\bibinfo{author}{\bibfnamefont{D.}~\bibnamefont{Budker}},
  \bibinfo{author}{\bibfnamefont{D.~F.} \bibnamefont{Kimball}},
  \bibinfo{author}{\bibfnamefont{S.~M.} \bibnamefont{Rochester}},
  \bibinfo{author}{\bibfnamefont{V.~V.} \bibnamefont{Yashchuk}},
  \bibnamefont{and}
  \bibinfo{author}{\bibfnamefont{M.}~\bibnamefont{Zolotorev}},
  \bibinfo{journal}{Phys. Rev. A} \textbf{\bibinfo{volume}{62}},
  \bibinfo{pages}{043403} (\bibinfo{year}{2000}).

\bibitem[{\citenamefont{Kominis et~al.}(2003)\citenamefont{Kominis, Kornack,
  Allred, and Romalis}}]{Kom2003}
\bibinfo{author}{\bibfnamefont{I.~K.} \bibnamefont{Kominis}},
  \bibinfo{author}{\bibfnamefont{T.~W.} \bibnamefont{Kornack}},
  \bibinfo{author}{\bibfnamefont{J.~C.} \bibnamefont{Allred}},
  \bibnamefont{and} \bibinfo{author}{\bibfnamefont{M.~V.}
  \bibnamefont{Romalis}}, \bibinfo{journal}{Nature}
  \textbf{\bibinfo{volume}{422}}, \bibinfo{pages}{596} (\bibinfo{year}{2003}).

\bibitem[{\citenamefont{Happer and Tam}(1977)}]{Hap77}
\bibinfo{author}{\bibfnamefont{W.}~\bibnamefont{Happer}} \bibnamefont{and}
  \bibinfo{author}{\bibfnamefont{A.~C.} \bibnamefont{Tam}},
  \bibinfo{journal}{Phys. Rev. A} \textbf{\bibinfo{volume}{16}},
  \bibinfo{pages}{1877} (\bibinfo{year}{1977}).

\bibitem[{\citenamefont{Budker et~al.}(2002{\natexlab{b}})\citenamefont{Budker,
  Kimball, Yashchuk, and Zolotorev}}]{Bud2002FM}
\bibinfo{author}{\bibfnamefont{D.}~\bibnamefont{Budker}},
  \bibinfo{author}{\bibfnamefont{D.~F.} \bibnamefont{Kimball}},
  \bibinfo{author}{\bibfnamefont{V.~V.} \bibnamefont{Yashchuk}},
  \bibnamefont{and}
  \bibinfo{author}{\bibfnamefont{M.}~\bibnamefont{Zolotorev}},
  \bibinfo{journal}{Phys. Rev. A} \textbf{\bibinfo{volume}{65}},
  \bibinfo{pages}{055403} (\bibinfo{year}{2002}{\natexlab{b}}).

\bibitem[{\citenamefont{Malakyan et~al.}(2004)\citenamefont{Malakyan,
  Rochester, Budker, Kimball, and Yashchuk}}]{Mal2004}
\bibinfo{author}{\bibfnamefont{Y.~P.} \bibnamefont{Malakyan}},
  \bibinfo{author}{\bibfnamefont{S.~M.} \bibnamefont{Rochester}},
  \bibinfo{author}{\bibfnamefont{D.}~\bibnamefont{Budker}},
  \bibinfo{author}{\bibfnamefont{D.~F.} \bibnamefont{Kimball}},
  \bibnamefont{and} \bibinfo{author}{\bibfnamefont{V.~V.}
  \bibnamefont{Yashchuk}}, \bibinfo{journal}{Phys. Rev. A}
  \textbf{\bibinfo{volume}{6901}}, \bibinfo{pages}{3817}
  (\bibinfo{year}{2004}).

\bibitem[{\citenamefont{Alexandrov et~al.}(2002)\citenamefont{Alexandrov,
  Balabas, Budker, English, Kimball, Li, and Yashchuk}}]{AleLIAD}
\bibinfo{author}{\bibfnamefont{E.~B.} \bibnamefont{Alexandrov}},
  \bibinfo{author}{\bibfnamefont{M.~V.} \bibnamefont{Balabas}},
  \bibinfo{author}{\bibfnamefont{D.}~\bibnamefont{Budker}},
  \bibinfo{author}{\bibfnamefont{D.}~\bibnamefont{English}},
  \bibinfo{author}{\bibfnamefont{D.~F.} \bibnamefont{Kimball}},
  \bibinfo{author}{\bibfnamefont{C.~H.} \bibnamefont{Li}}, \bibnamefont{and}
  \bibinfo{author}{\bibfnamefont{V.~V.} \bibnamefont{Yashchuk}},
  \bibinfo{journal}{Phys. Rev. A} \textbf{\bibinfo{volume}{66}},
  \bibinfo{pages}{042903/1} (\bibinfo{year}{2002}).

\bibitem[{\citenamefont{Yashchuk and Budker}(2004)}]{Yas2004ModDAVLL}
\bibinfo{author}{\bibfnamefont{V.~V.} \bibnamefont{Yashchuk}} \bibnamefont{and}
  \bibinfo{author}{\bibfnamefont{D.}~\bibnamefont{Budker}},
  \emph{\bibinfo{title}{To be published}} (\bibinfo{year}{2004}).

\bibitem[{\citenamefont{Yashchuk et~al.}(1999)\citenamefont{Yashchuk, Budker,
  and Zolotorev}}]{Yas99}
\bibinfo{author}{\bibfnamefont{V.}~\bibnamefont{Yashchuk}},
  \bibinfo{author}{\bibfnamefont{D.}~\bibnamefont{Budker}}, \bibnamefont{and}
  \bibinfo{author}{\bibfnamefont{M.}~\bibnamefont{Zolotorev}}, in
  \emph{\bibinfo{booktitle}{Trapped Charged Particles and Fundamental Physics}}
  (\bibinfo{address}{Asilomar, CA, USA}, \bibinfo{year}{1999}), vol.
  \bibinfo{volume}{457} of \emph{\bibinfo{series}{AIP Conf. Proc.}}, pp.
  \bibinfo{pages}{177--81}.

\bibitem[{\citenamefont{Yashchuk et~al.}(2004)\citenamefont{Yashchuk, Granwehr,
  Trabesinger, and Budker}}]{Yas2004NMRtube}
\bibinfo{author}{\bibfnamefont{V.~V.} \bibnamefont{Yashchuk}},
  \bibinfo{author}{\bibfnamefont{J.}~\bibnamefont{Granwehr}},
  \bibinfo{author}{\bibfnamefont{A.~H.} \bibnamefont{Trabesinger}},
  \bibnamefont{and} \bibinfo{author}{\bibfnamefont{D.}~\bibnamefont{Budker}},
  \emph{\bibinfo{title}{To be published}} (\bibinfo{year}{2004}).

\bibitem[{\citenamefont{Haake et~al.}(1998)\citenamefont{Haake, Goodson, Laws,
  Brunner, Cyrier, Havlin, and Pines}}]{Haa98}
\bibinfo{author}{\bibfnamefont{M.}~\bibnamefont{Haake}},
  \bibinfo{author}{\bibfnamefont{B.~M.} \bibnamefont{Goodson}},
  \bibinfo{author}{\bibfnamefont{D.~D.} \bibnamefont{Laws}},
  \bibinfo{author}{\bibfnamefont{E.}~\bibnamefont{Brunner}},
  \bibinfo{author}{\bibfnamefont{M.~C.} \bibnamefont{Cyrier}},
  \bibinfo{author}{\bibfnamefont{R.~H.} \bibnamefont{Havlin}},
  \bibnamefont{and} \bibinfo{author}{\bibfnamefont{A.}~\bibnamefont{Pines}},
  \bibinfo{journal}{Chem. Phys. Lett.} \textbf{\bibinfo{volume}{292}},
  \bibinfo{pages}{686} (\bibinfo{year}{1998}).

\end{thebibliography}

\end{document}